\def\be{\begin{equation}}
\def\ee{\end{equation}}
\def\ba{\begin{array}}
\def\ea{\end{array}}

\def\Cb{{\Bbb C}}

\documentclass[preprint,showpacs,amsmath]{revtex4}
\usepackage{amsfonts}
\usepackage{amssymb}
\def\qed{\leavevmode\unskip\penalty9999 \hbox{}\nobreak\hfill
     \quad\hbox{\leavevmode  \hbox to.77778em{%
               \hfil\vrule   \vbox to.675em%
               {\hrule width.6em\vfil\hrule}\vrule\hfil}}
     \par\vskip3pt}

\input amssym.def
\begin{document}
\begin{center}\bf\large
{Criterion of Local Unitary Equivalence for Multipartite States}
\end{center}
\vskip 1mm
\begin{center}

{Ting-Gui Zhang$^{1}$}, {Ming-Jing Zhao$^{2}$}, {Ming Li}$^{1,3}$,
 {Shao-Ming Fei$^{1,4}$} and {Xianqing Li-Jost$^{1}$}\vspace{2ex}

\begin{minipage}{5.5in}
{\small $^1$Max-Planck-Institute for Mathematics in the Sciences, 04103 Leipzig, Germany\\
$^2$Department of Mathematics, School of Science, Beijing
Information Science and
Technology University, 100192, Beijing, China\\
$^3$Department of Mathematics, School of Science, China University
of Petroleum, 266555 Qingdao, China\\
$^4$School of Mathematical
Sciences, Capital Normal University, Beijing 100048, China}
\end{minipage}

\end{center}

\begin{abstract}

We study the local unitary equivalence of arbitrary dimensional
multipartite quantum mixed states. We present a necessary and
sufficient criterion of the local unitary equivalence for general
multipartite states based on matrix realignment. The criterion is
shown to be operational even for particularly degenerated states by
detailed examples. Besides, explicit expressions of the local
unitary operators are constructed for locally equivalent states. In
complement to the criterion, an alternative approach based on
partial transposition of matrices is also given, which makes the
criterion more effective in dealing with generally degenerated mixed
states.
\end{abstract}

 \maketitle

\maketitle

\bigskip

Quantum entanglement is one of the most extraordinary features of quantum
physics. Multipartite entanglement plays a vital role in quantum information
processing \cite{bdfk,EofCon} and interferometry \cite{nils}.
One fact is that the degree of entanglement of a quantum state remains invariant under local
unitary transformations, while two quantum states with the same
degree entanglement, e.g. entanglement of formation \cite{Cbben,
mhor} or concurrence \cite{Auhl, pvcm}), may be not equivalent under
local unitary transformations.
Another fact is that two entangled states are said to be equivalent in
implementing quantum information tasks, if they can be
mutually exchanged under local operations and classical
communication (LOCC). LOCC equivalent states are
interconvertible also by local unitary transformations \cite{Wdur}.
Therefore, it is important to classify and characterize quantum states in terms of local unitary
transformations.

To deal with this problem, one approach is to construct invariants
of local unitary transformations. The method developed in
\cite{Rains,Grassl}, in principle, allows one to compute all the
invariants of local unitary transformations for bipartite states,
though it is not easy to do this operationally. In \cite{makhlin} a
complete set of 18 polynomial invariants is presented for the local
unitary equivalence of two qubits mixed states. Partial results have
been obtained for three qubits states \cite{linden, Linden}, some
generic mixed states \cite{SFG, SFW, SFY}, tripartite pure and mixed
states \cite{SCFW}. In \cite{sss}  give explicit index-free formulae
for all the degree $6$ algebraically independent local unitary
invariant polynomials for finite-dimensional k-partite pure and
mixed quantum states. The local unitary equivalence problem for
multipartite pure qubits states has been solved in \cite{mqubit}. By
exploiting the high order singular value decomposition technique and
local symmetries of the states, Ref. \cite{bliu} presents a
practical scheme of classification under local unitary
transformations for general multipartite pure states with arbitrary
dimensions, which extends results of n-qubit pure states
\cite{mqubit} to that of n-qudit pure states. For mixed states, Ref.
\cite{zhou} solved the local unitary equivalence problem of
arbitrary dimensional bipartite non-degenerated quantum systems by
presenting a complete set of invariants, such that two density
matrices are local unitary equivalent if and only if all these
invariants have equal values. In \cite{zhang} the case of
multipartite systems is studied and a complete set of invariants is
presented for a special class of mixed states.

In this paper, we study the local unitary equivalence problem in
terms of matrix realignment \cite{o.ru, Chen} and partial
transposition \cite{AP, MH}, the techniques used in dealing with the
separability problem of quantum states and also in generating local
unitary invariants \cite{bsl}. We present a necessary and sufficient
criterion for the local unitary equivalence of multipartite states,
together with explicit forms of the local unitary operators. This
generalizes the results in \cite{zhou,feij} from non-degenerated
states to generally degenerated states for bipartite case. The
criterion is shown to be still operational for states having
eigenvalues with multiplicity no more than $2$. It also generalizes
the results in \cite{zhou,feij} from bipartite states to generally
multipartite states. Alternative ways are presented to deal with
generally degenerated states by using our criterion.

We first review some definitions and results about matrix
realignment from matrix analysis \cite{Raho}. For any $M\times N$
matrix $A$ with entries $a_{ij}$, $vec(A)$ is defined by
$$
vec(A)\equiv[a_{11},\cdots,a_{M1},a_{12}\cdots,a_{M2},\cdots,a_{1N},\cdots,a_{MN}]^T,
$$
where $T$ denotes transposition. Let $Z$ be an $M\times M$ block
matrix with each block of size $N\times N$, the realigned matrix
$\widetilde{Z}$ is defined by
$$
\widetilde{Z}\equiv[vec(Z_{11}),\cdots,vec(Z_{M1}),\cdots,vec(Z_{1M}),\cdots,vec(Z_{MM})]^T.
$$

Based on these operations, the authors in \cite{npp,cfvl} proved that

\noindent {\bf Lemma 1:} Assume that the matrix $\widetilde{Z}$ has
singular value decomposition, $\widetilde{Z}=U\Sigma V^\dag$, then
$Z=\sum_{i=1}^rX_i\otimes Y_i$, where
$vec(X_i)=\sqrt{\alpha_i\sigma_i}\mu_i$,
$vec(Y_i)=\sqrt{\frac{1}{\alpha_i}\sigma_i}\nu_i^\ast$,
$\alpha_i\neq 0$, $\Sigma=diag(\sigma_i)$ with $\sigma_1 \geq
\sigma_2 \geq \cdots \geq \sigma_q \geq 0$, $\{\sigma_i\}_{i=1}^q$
are the singular values of the matrix $\widetilde{Z}$,
$q=min(M^2,N^2)$, $r$ is the number of nonzero singular values
$\sigma_i$ (the rank of the matrix $\widetilde{Z}$),
$U=[\mu_1\mu_2\cdots \mu_{M^2}]\in \mathbb{C}^{M^2\times M^2}$ and
$V=[\nu_1\nu_2\cdots \nu_{N^2}]\in \mathbb{C}^{N^2\times N^2}$ are
unitary matrices, with $\mu_i$ and $\nu_i$ the singular vectors of
$\sigma_i$.

Lemma 1 implies that \cite{slsx},

\noindent {\bf Lemma 2} An $MN\times MN$ unitary
matrix $U$ can be expressed as the
tensor product of an $M\times M$ unitary matrix
$u_1$ and an $N\times N$ unitary matrix $u_2$ such that $U=u_1\otimes u_2$ if and only if rank
$(\widetilde{U})=1$.

{\sf Remark 1}: \ Following Lemma 1, when rank$(\widetilde{U})=1$,
$vec(X)=\sqrt{\alpha_1\sigma_1}\mu_1$ and
$vec(Y)=\sqrt{\frac{1}{\alpha_1}\sigma_1}\nu_1^\ast$, where $\mu_1$
and $\nu_1$ are the eigenvectors of
$\widetilde{U}\widetilde{U}^{\dag}$ and
$\widetilde{U}^{\dag}\widetilde{U}$ corresponding to non-zero
eigenvalues. Therefore, from Lemma 2, the detailed form of $u_1$ and
$u_2$ can be obtained.

Now consider the case of multipartite states. Let $H_1,
H_2,\cdots,H_n$ be complex Hilbert spaces of finite
dimensions $N_1, N_2,$ $\cdots,N_n$, respectively. Let
$\{|j\rangle_k\}_{j=1}^{N_k}$, $k = 1,2,\cdots,n$, be an orthonormal
basis of $H_k$. A mixed state $\rho\in H_1\otimes H_2\otimes
\cdots\otimes H_n$ can be written in terms of the spectral decomposition form of $\rho$,
$\rho=\sum_{i=1}^{N_1N_2\cdots
N_n}\lambda_i|\phi_i\rangle\langle\phi_i|$, where
$|\phi_i\rangle=\sum_{j=1}^{N_1}\sum_{k=1}^{N_2}\cdots\sum_{l=1}^{N_n}a_{jk\cdots
l}^i|j\rangle_1|k\rangle_2\cdots|l\rangle_n$, $a_{jk\cdots l}^i\in\Cb$ satisfying
$\sum_{j=1}^{N_1}\sum_{k=1}^{N_2}\cdots\sum_{l=1}^{N_n}a_{jk\cdots
l}^ia_{jk\cdots l}^{i\dag}=1$.

Two multipartite mixed states $\rho$ and $\rho^\prime$ in
$H_1\otimes H_2\otimes \cdots \otimes H_n$ are said to be equivalent
under local unitary transformations if there exist unitary operators
$u_i$ on the $i$-th Hilbert space $H_i$ such that \be\label{eq}
\rho^\prime=(u_1\otimes u_2\otimes\cdots \otimes u_n)\rho(u_1\otimes
u_2\otimes \cdots \otimes u_n)^\dag. \ee

In the following, for any $N_1N_2\cdots N_n\times N_1N_2\cdots N_n$
matrix $T$, we denote $T_{i|\widehat{i}}$ the $N_i\times N_i$ block
matrix with each block of size $N_1N_2\cdots N_{i-1} N_{i+1}\cdots
N_n\times N_1N_2\cdots N_{i-1} N_{i+1}\cdots N_n$. Namely, we view
$T$ as a bipartite partitioned matrix $T_{i|\widehat{i}}$ with
partitions $H_i$ and $H_1\otimes H_2 ... H_{i-1}\otimes
H_{i+1}...H_n$. Accordingly, we have the realigned matrix
$\widetilde{T_{i|\widehat{i}}}$.

\noindent {\bf Lemma 3} Let $U$ be an $N_1N_2\cdots N_n\times
N_1N_2\cdots N_n$ unitary matrix, there exist $N_i\times N_i$
unitary matrices $u_i$, $i=1,2,\cdots, n$, such that $U=u_1\otimes
u_2 \otimes \cdots\otimes u_n$ if and only if the
rank$(\widetilde{U_{i|\widehat{i}}})=1$ for all $i$.

\noindent {\bf Proof}\ \ First, if there exist $N_i\times N_i$
 unitary matrices
$u_i$, $i=1,2,\cdots, n$, such that $U=u_1\otimes u_2 \otimes \cdots
\otimes u_n$, by viewing $U$ in bipartite partition and using Lemma
2, one has directly that rank$(\widetilde{U_{i|\widehat{i}}})=1$ for
all $i$.

On the other hand, if rank$(\widetilde{U_{i|\widehat{i}}})=1$, for
any given $i$, we prove the conclusion by induction. First, for
$n=3$, from Lemma 2, we have $U=u_1\otimes u_{23}=u_2\otimes
u_{13}$, i.e, $(u_1^{\dag}\otimes I_{2}\otimes I_{3})U=I_1\otimes
u_{23}=u_2\otimes ((u_1^{\dag}\otimes I_3)u_{13})$. By tracing over
the first subsystem, we get $N_1u_{23}=u_2\otimes
Tr_1((u_1^{\dag}\otimes I_3)u_{13})$, i.e, $u_{23}=u_2\otimes
u_3^{\prime}$ with $u_3^{\prime}=Tr_1((u_1^{\dag}\otimes
I_3)u_{13})/N_1$. Assume that the conclusion is also true for $n-1$.
Then for $n$, from Lemma 2, we have $U=u_1\otimes
u_{\widehat{1}}=u_2\otimes u_{\widehat{2}}=\cdots=u_n\otimes
u_{\widehat{n}}$, where $u_i$ is an $N_i\times N_i$ unitary matrix
and $u_{\widehat{i}}$ is an $N_1 N_2\cdots N_{i-1} N_{i+1}\cdots
N_n\times N_1 N_2\cdots N_{i-1} N_{i+1}\cdots N_n$ unitary matrix,
$i=1,2,\cdots,n$. Hence $(I_1\otimes\cdots \otimes I_{n-1}\otimes
u_n^{\dag})U=(I_1\otimes\cdots \otimes I_{n-1}\otimes
u_n^{\dag})(u_1\otimes
u_{\widehat{1}})=\cdots=u_{\widehat{n}}\otimes I_{N_n}$. By tracing
the last subsystem we get $u_1\otimes (Tr_n(I_2\otimes\cdots \otimes
I_{N_{n-1}}\otimes
u_n^{\dag})u_{\widehat{1}})=\cdots=(Tr_n(I_1\otimes\cdots \otimes
I_{n-2}\otimes u_n^{\dag}))\otimes(u_{n-1})=N_nu_{\widehat{n}}$.
Based on the assumption, we have that $u_{\widehat{n}}$ can be
written as the tensor of local unitary operators. Therefore, $U$
also can be written as the tensor product of local unitary
operators. \qed

If two density matrices $\rho_1$ and $\rho_2$ in $H_1 \otimes
H_2\otimes \cdots \otimes H_n$ are equivalent under local unitary
transformations, they must have the same set of eigenvalues
$\lambda_k, \ k=1,2,\cdots, N_1N_2\cdots N_n$. Let
$X=(x_1,x_2,\cdots,x_{N_1N_2\cdots N_n})$ and
$Y=(y_1,y_2,\cdots,y_{N_1N_2\cdots N_n})$ be the unitary matrices
that diagonalize the two density matrices, respectively,
\be\label{eq1}
\rho_1=X\Lambda X^\dag, \ \ \rho_2=Y\Lambda Y^\dag,
\ee
where $\{x_i\}$ and $\{y_i\}$ are the normalized eigenvectors of
states $\rho_1$ and $\rho_2$,
$$
\Lambda=diag( \lambda_1I_{n_1}, \lambda_2I_{n_2}, \cdots,
\lambda_rI_{n_r}),
$$
with $r \leq N_1N_2\cdots N_n,\  \sum_{k=1}^rn_k=N_1N_2\cdots N_n$,
$n_k$ is the multiplicity of the $k$th eigenvalue $\lambda_k$.
Therefore $X^{\dag}\rho_1 X=\Lambda=Y^{\dag}\rho_2 Y$.
Due to the degeneracy of $\rho_1$ and $\rho_2$,
$X$ and $Y$ are not fixed in the sense that
$X^{\dag}\rho_1 X=Y^{\dag}\rho_2 Y$ is inveriant under
$X\to XU$ and $Y\to YU$, for any
\be \label{eq2}
U=diag( u_{n_1},
 u_{n_2}, \cdots, u_{n_r} ),
\ee
where $u_{n_k}$ are $n_k\times n_k$ unitary
matrices, $k=1,\cdots,r$. Thus for given $X$ and $Y$, $YU^\dag
X^{\dag}\rho_1 XUY^{\dag}=\rho_2$.

\noindent {\bf Theorem 1}~ Let $\rho_1$ and $\rho_2$ be two
multipartite mixed quantum states given in (\ref{eq1}),
$\rho_1=X\Lambda X^\dag$ and $\rho_2=Y\Lambda Y^\dag$. $\rho_1$ and
$\rho_2$ are local unitary equivalent if and only if there exists an
$N_1N_2\cdots N_n\times N_1N_2\cdots N_n$ unitary matrix $U$ of the
form (\ref{eq2}) such that
rank$(\widetilde{XUY^{\dag}})_{i|\widehat{i}}=1$ for $i= 1,2,\cdots,
n$.

\noindent {\bf Proof:}~ If $\rho_1$ and $\rho_2$ are equivalent
under local unitary transformations, i.e. $(u_1\otimes
u_2\otimes\cdots \otimes u_n)\rho_1(u_1\otimes
u_2\otimes\cdots\otimes u_n)^\dag=\rho_2$, then there exists a
unitary matrix $U$ of the form (\ref{eq2}) such that $Y=(u_1\otimes
u_2\otimes\cdots\otimes u_n)XU$. From Lemma 3 the
rank$(\widetilde{XUY^{\dag}})_{i|\widehat{i}}=1$,
where $(XUY^{\dag})_{i|\widehat{i}}={u_i\otimes
(u_1\otimes\cdots\otimes u_{i-1}\otimes u_{i+1}\otimes\cdots\otimes
u_n)}$, $i=1,2,\cdots,n$.

On the other hand, if there is an $N_1N_2\cdots N_n\times
N_1N_2\cdots N_n$ unitary matrix $U$ such that
rank$(\widetilde{XUY^{\dag}})_{i|\widehat{i}}=1$, for any $i$, by
Lemma 3 we have $XU Y^{\dag}=u_1\otimes u_2\otimes \cdots \otimes
u_n$. Then $YU^\dag X^{\dag}\rho_1 XUY^{\dag}=\rho_2$ gives rise to
$(u_1\otimes u_2 \otimes\cdots\otimes u_n)^\dag\rho_1(u_1\otimes u_2
\otimes \cdots\otimes u_n)=\rho_2$, which ends the proof.\qed

{\sf Remark 2}: \ If there exists an $N_1N_2\cdots N_n\times
N_1N_2\cdots N_n$ unitary matrix $U$ of the form (\ref{eq2}) such
that rank$(\widetilde{XUY^{\dag}})_{i|\widehat{i}}=1$, for any $i=
1,2,\cdots, n$, then $\rho_1$ and $\rho_2$ are local unitary
equivalent. From Lemma 1, we can get the explicit expressions of the
local unitary matrices $u_i$ in the following way. First, we view
$XUY^\dag$ as an $N_1\times N_1$ block matrix with each block of
size $N_2N_3\cdots N_n\times N_2N_3\cdots N_n$. Following Lemma 1,
we have that $XUY^\dag=u_1\otimes u_{\widehat{1}}$, where $u_1$ and
$u_{\widehat{1}}$ have explicit expressions from Remark 1, and
$u_{\widehat{1}}$ is an $N_2N_3\cdots N_n\times N_2N_3\cdots N_n$
unitary matrix. By viewing $u_{\widehat{1}}$ as an $N_2\times N_2$
block matrix with each block of size $N_3N_4\cdots N_n\times
N_3N_4\cdots N_n$, we get the expression of $u_2$ in
$u_{\widehat{1}}=u_2\otimes u_{\widehat{2}}$. In this way, we can
get all the detailed expressions of $u_1,u_2,\cdots,u_n$, such that
$XUY^\dag=u_1\otimes u_2\otimes\cdots\otimes u_n$.

Theorem 1 has many advantages compared with the previous results
about local unitary equivalence. It generalizes the results for
non-degenerated bipartite states in \cite{zhou} to general bipartite
mixed states including degenerated ones, for which the problem
becomes quite difficult usually and many criteria become
non-operational \cite{zhou}. Our criterion can be also operational
for particular degenerated bipartite states. Let us consider that
$\rho_1,\rho_2\in H_1\otimes H_2$ have $s$ different eigenvalues
with multiplicity $2$ and the rest eigenvalues with multiplicity
$1$. According to Theorem 1, $\rho_1$ and $\rho_2$ are local unitary
equivalent if and only if there exists a unitary matrix
\be\label{eq3}
U=diag(u_1,\cdots,u_s,e^{i\theta_{s+1}},\cdots,e^{i\theta_{N_1N_2}}),
\ee with $u_r\in U(2)$, $r=1,\cdots, s$,
$s=0,1,\cdots,[\frac{N_1N_2}{2}]$, such that
rank$(\widetilde{XUY^{\dag}})=1$, where $[x]$ denotes the integer
part of $x$.

Any unitary matrix in $U(2)$ can be written as, up to a constant
phase, $tI+ i \sum_{j=1}^3{z}_j{\sigma}_j$ with $t^2+\sum_{j=1}^3
z_j^2=1$, where $I$ is the $2\times 2$ identity matrix and
$\sigma_j$ are the Pauli matrices. Therefore $U$ has the following
form: \be\label{U} U=\left(
\begin{array}{ccccccccc}
t_1+iz_3 & z_1+iz_2 & \cdots &0 & 0 & 0 & \cdots & 0 & 0\\
-z_1+iz_2 & t_1-iz_3 & \cdots & 0 & 0 & 0 & \cdots & 0 &0\\
\cdots & \cdots & \cdots & \cdots &\cdots & \cdots& \cdots & \cdots & \cdots\\
0 & 0 & \cdots & t_s+iz_{3s} & z_{3s-2}+iz_{3s-1}& 0 &\cdots & 0 & 0 \\
0 & 0 & \cdots & -z_{3s-2}+iz_{3s-1} & t_{s}-iz_{3s} & 0 &\cdots & 0 & 0 \\
0& 0& \cdots &0 & 0 & e^{i\theta_{s+1}}  &\cdots & 0 & 0 \\
\cdots& \cdots& \cdots &\cdots & \cdots & \cdots &\cdots & \cdots & \cdots \\
0& 0& \cdots &0 & 0 & \cdots &\cdots & 0  & e^{i\theta_{N_1N_2}}
\end{array}
\right), \ee where $t^2_j+ z_{3j}^2 + z_{3j-1}^2 + z_{3j-2}^2=1$ for
$j=1,...,s$. One just needs to verify the existence of the unitary
matrix $U$ such that rank$(\widetilde{XUY^{\dag}})=1$. The
calculation of the rank of $\widetilde{XUY}$ only concerns the
quadratic homogeneous equations and can be done simply by using the
algorithm in Ref. \cite{naja} for solving systems of multivariate
polynomial equations called XL (eXtended Linearizations or
multiplication and linearlization) algorithm.

\noindent {\bf Example 1:} As an example, let us consider two bipartite states:
$$\rho_1=\left(\begin{array}{cccc}
1/4 & 0 & 0 & 1/4 \\
0 & 1/4 & 1/4 & 0 \\
0 & 1/4 & 1/4 & 0 \\
1/4 & 0 & 0 & 1/4
\end{array}
\right),~~~~ \rho_2=\left(\begin{array}{cccc}
1/4 & 0 & 1/4 & 0 \\
0 & 1/4 & 0 & -1/4 \\
1/4 & 0 & 1/4 & 0 \\
0 & -1/4 & 0 & 1/4
\end{array}
\right).
$$
Here $\rho_1$ and $\rho_2$ are degenerated states with the
eigenvalues set $\Lambda=diag(\frac{1}{2},\frac{1}{2},0,0)$.
They can be written as,
$\rho_{1}=\frac{1}{2}|\phi_{1}\rangle\langle\phi_{1}|+\frac{1}{2}|\phi_{2}\rangle\langle\phi_{2}|$,
$\rho_{2}=\frac{1}{2}|\varphi_{1}\rangle\langle\varphi_{1}|+\frac{1}{2}|\varphi_{2}\rangle\langle\varphi_{2}|$,
where $|\phi_{1}\rangle=\frac{1}{\sqrt{2}}(|00\rangle+|11\rangle)$,
$|\phi_{2}\rangle=\frac{1}{\sqrt{2}}(|01\rangle+|10\rangle)$,
$|\varphi_{1}\rangle=-\frac{1}{\sqrt{2}}[(|0\rangle-|1\rangle)\otimes
|1\rangle]$, and
$|\varphi_{2}\rangle=\frac{1}{\sqrt{2}}[(|0\rangle+|1\rangle)\otimes
|0\rangle]$. One can prove that $\rho_{1}$ and $\rho_{2}$ do not
satisfy the following conditions,
$$
Tr[(A_{\alpha}A_{\alpha}^{\dag})^{p}]=Tr[(B_{\alpha}B_{\alpha}^{\dag})^{p}],\
\  \ \alpha=1,2,~~ \ \ \ \ p=1,2,3,4,
$$
where $A_\alpha$ (resp. $B_\alpha$) denotes the matrix representation of the bipartite pure
state $|\phi_{\alpha}\rangle$ (resp. $|\varphi_{\alpha}\rangle$),
$\alpha=1,2$ \cite{zhou}. Therefore, the criterion given in
\cite{zhou} fails to detect the local unitary equivalence of
$\rho_1$ and $\rho_2$.

Following (\ref{eq2}), $U$ has the form
$$
U=\left(
\begin{array}{cccc}
t_1+iz_3 & z_1+iz_2 & 0 & 0 \\
-z_1+iz_2 & t_1-iz_3 &  0 & 0 \\
0 & 0 &  t_2+iz_{6} & z_{4}+iz_{5} \\
0 & 0 & -z_{4}+iz_{5} & t_{2}-iz_{6}
\end{array}
\right).
$$
Correspondingly,
$$X=\left(\begin{array}{cccc}
1/\sqrt{2} & 0 & -1/\sqrt{2} & 0 \\
0 & 1/\sqrt{2} & 0 & -1/\sqrt{2} \\
0 & 1/\sqrt{2} & 0 & 1/\sqrt{2} \\
1/\sqrt{2} & 0 & 1/\sqrt{2} & 0
\end{array}
\right),~~~ Y=\left(\begin{array}{cccc}
1/\sqrt{2} & 0 & -1/\sqrt{2} & 0 \\
0 & -1/\sqrt{2} & 0 & 1/\sqrt{2} \\
1/\sqrt{2} & 0 & 1/\sqrt{2} & 0 \\
0 & 1/\sqrt{2} & 0 & 1/\sqrt{2}
\end{array}
\right).
$$
It is easily verified that there are many matrices of the form
(\ref{U}) satisfying rank$(\widetilde{XUY^\dag})=1$, for instance,
$$
U=\left(\begin{array}{cccc}
-\frac{1}{\sqrt{2}}& \frac{1}{\sqrt{2}} & 0 & 0 \\
-\frac{1}{\sqrt{2}} & -\frac{1}{\sqrt{2}} & 0 & 0 \\
0 & 0 & -\frac{1}{\sqrt{2}} & \frac{1}{\sqrt{2}}\\
0 & 0 & -\frac{1}{\sqrt{2}} &-\frac{1}{\sqrt{2}}
\end{array}
\right).
$$
Therefore $\rho_1$ and $\rho_2$ are local unitary equivalent. In
fact, from singular values decomposition of
$\widetilde{XUY^{\dag}}$, we can get the unique nonzero singular
values $\frac{1}{2}$ with multiplicity 2. Using Lemma 1, we have
$\mu_1=\frac{1}{\sqrt{2}}(-1,0,0,1)$ and $\nu_1=\frac{1}{2}(1, -1,
1, 1)$. Therefore, from Lemma 2, we can choose
$vec(X_1)=\sqrt{2}u_1$ and $vec(Y_1)=\sqrt{2}v_1$, such that $X_1$
and $Y_1$ are unitary matrices, and $(X_1\otimes
Y_1)\rho_1(X_1\otimes Y_1)^\dag=\rho_2$.

Now, we consider multipartite case. For multipartite pure states,
our approach has the same operational difficulty as the previous ones in
\cite{bliu}. For multipartite mixed states, in particular,
for non-degenerated states, our criterion can give effective
procedures to detect the local unitary equivalence of two given
states.

\noindent {\bf Example 2:} Consider two density matrices in
$H_1\otimes H_2 \otimes H_3$ with $N_1=N_2=N_3=2$,
$$
\rho_1=\frac{1}{K}\left(\begin{array}{cccccccc}
1 & 0 & 0 & 0 & 0 & 0 & 0 & 1\\
0 & a & 0 & 0 & 0 & 0 & 0 & 0\\
0 & 0 & b & 0 & 0 & 0 & 0 & 0\\
0 & 0 & 0 & c & 0 & 0 & 0 & 0\\
0 & 0 & 0 & 0 & \frac{1}{c} & 0 & 0 & 0\\
0 & 0 & 0 & 0 & 0 & \frac{1}{b} & 0 & 0\\
0 & 0 & 0 & 0 & 0 & 0 & \frac{1}{a} & 0\\
1 & 0 & 0 & 0 & 0 & 0 & 0 & 1
\end{array}
\right),~~~\rho_2=\frac{1}{K}\left(\begin{array}{cccccccc}
\frac{1+b}{2} & 0 & \frac{b-1}{2} & 0 & 0 & \frac{1}{2} & 0 & \frac{1}{2}\\
0 & \frac{a+c}{2} & 0 & \frac{c-a}{2} & 0 & 0 & 0 & 0\\
\frac{b-1}{2} & 0 & \frac{1+b}{2} & 0 & 0 & -\frac{1}{2} & 0 & \frac{1}{2}\\
0 & \frac{c-a}{2} & 0 & \frac{a+c}{2} & 0 & 0 & 0 & 0\\
0 & 0 & 0 & 0 & \frac{1}{2c}+\frac{1}{2a} & 0 & \frac{1}{2a}-\frac{1}{2c} & 0\\
\frac{1}{2} & 0 & -\frac{1}{2} & 0 & 0 & \frac{1}{2b}+\frac{1}{2} & 0 & \frac{1}{2}-\frac{1}{2b}\\
0 & 0 & 0 & 0 & \frac{1}{2a}-\frac{1}{2c} & 0 & \frac{1}{2c}+\frac{1}{2a} & 0\\
\frac{1}{2} & 0 & -\frac{1}{2} & 0 & 0 & \frac{1}{2}-\frac{1}{2b} &
0 & \frac{1}{2}+\frac{1}{2b}
\end{array}
\right),
$$
where the normalization factor
$K=2+a+b+c+\frac{1}{a}+\frac{1}{b}+\frac{1}{c}$. $\rho_1$ and
$\rho_2$ have the same eigenvalue set $\Lambda= \frac{1}{K} diag
(2,0,\frac{1}{a},a,\frac{1}{b},b,\frac{1}{c},c)$. For the case
$a\neq b\neq c\neq 0\neq 1\neq 2\neq \frac{1}{2}$, $\rho_1$ and
$\rho_2$ are not degenerated. In this case, one has
$$
X=\left(\begin{array}{cccccccc}
\frac{1}{\sqrt{2}} & -\frac{1}{\sqrt{2}}& 0 & 0 & 0 & 0 & 0 & 0\\
0 & 0 & 0 & 1 & 0 & 0 & 0 & 0\\
0 & 0 & 0 & 0 & 0 & 1 & 0 & 0\\
0 & 0 & 0 & 0 & 0 & 0 & 0 & 1\\
0 & 0 & 0 & 0 & 0 & 0 & 1 & 0\\
0 & 0 & 0 & 0 & 1 & 0 & 0 & 0\\
0 & 0 & 1 & 0 & 0 & 0 & 0 & 0\\
\frac{1}{\sqrt{2}} & \frac{1}{\sqrt{2}} & 0 & 0 & 0 & 0 & 0 & 0
\end{array}
\right),~~~
Y=\left(\begin{array}{cccccccc}
\frac{1}{2} & -\frac{1}{2}& 0 & 0 & 0 & \frac{1}{\sqrt{2}} & 0 & 0\\
0 & 0 & 0 & -\frac{1}{\sqrt{2}} & 0 & 0 & 0 & \frac{1}{\sqrt{2}}\\
-\frac{1}{2} & \frac{1}{2} & 0 & 0 & 0 & \frac{1}{\sqrt{2}} & 0 & 0\\
0 & 0 & 0 & \frac{1}{\sqrt{2}} & 0 & 0 & 0 & \frac{1}{\sqrt{2}}\\
0 & 0 & \frac{1}{\sqrt{2}} & 0 & 0 & 0 & -\frac{1}{\sqrt{2}} & 0\\
\frac{1}{2} & \frac{1}{2} & 0 & 0 & -\frac{1}{\sqrt{2}} & 0 & 0 & 0\\
0 & 0 & \frac{1}{\sqrt{2}} & 0 & 0 & 0 & \frac{1}{\sqrt{2}} & 0\\
\frac{1}{2} & \frac{1}{2} & 0 & 0 & \frac{1}{\sqrt{2}} & 0 & 0 & 0
\end{array}
\right).
$$
From (\ref{eq2}), $U$ is of the form
$U=diag(e^{i\theta_1},e^{i\theta_2},e^{i\theta_3},e^{i\theta_4},e^{i\theta_5},e^{i\theta_6},e^{i\theta_7},e^{i\theta_8})$.
Hence
$$
XUY^{\dag}=\frac{1}{\sqrt{2}}\left(\begin{array}{cccccccc}
\frac{e^{i\theta_1}+e^{i\theta_2}}{2} & 0 &
-\frac{e^{i\theta_1}+e^{i\theta_2}}{2} & 0 & 0
& \frac{e^{i\theta_1}-e^{i\theta_2}}{2} & 0 & \frac{e^{i\theta_1}-e^{i\theta_2}}{2}\\
0 & -e^{i\theta_4} & 0 & e^{i\theta_4} & 0 & 0 & 0 & 0 \\
e^{i\theta_6} & 0 & e^{i\theta_6} & 0 & 0 & 0 & 0 & 0\\
0 & e^{i\theta_8} & 0 & e^{i\theta_8} & 0 & 0 & 0 & 0\\
0 & 0 & 0 & 0 &-e^{i\theta_7} & 0 & e^{i\theta_7} & 0\\
0 & 0 & 0 & 0 & 0 & -e^{i\theta_5} & 0 & e^{i\theta_5}\\
0 & 0 & 0 & 0 & e^{i\theta_3} & 0 & e^{i\theta_3} & 0\\
\frac{e^{i\theta_1}-e^{i\theta_2}}{2} & 0 &
\frac{e^{i\theta_2}-e^{i\theta_1}}{2} & 0 & 0 &
\frac{e^{i\theta_1}+e^{i\theta_2}}{2} & 0 &
\frac{e^{i\theta_1}+e^{i\theta_2}}{2}
\end{array}
\right).
$$
It is easily verified that
rank$(\widetilde{XUY^{\dag}})_{i|\widehat{i}}=1$ for
$\theta_1=\theta_2=\theta_3=\theta_6=\theta_8=0,
\theta_4=\theta_5=\theta_7=\pi$, $i=1,2,3$. Therefore from Theorem 1
$\rho_1$ and $\rho_2$ are local unitary equivalent. In fact, taking
$i=1$, from the singular values decomposition of
$(\widetilde{XUY^{\dag}})_{1|\widehat{1}}$, we can get the unique
nonzero singular values $2\sqrt{2}$. From Lemma 1, we get
$u_1=\frac{1}{\sqrt{2}}(1,0,0,1)$ and $v_1=\frac{1}{2\sqrt{2}}(1, 0,
1, 0, 0, 1, 0, 1, -1, 0, 1, 0, 0, -1, 0, 1)$. Therefore, we can
choose $vec(X_1)=\sqrt{2}u_1$ and $vec(X_2)=2v_1$ such that they are
unitary. Then $X_1=I_2\in H_1$ and
$$
X_2=\frac{1}{\sqrt{2}}\left(\begin{array}{cccc}
1 & 0 & -1 & 0 \\
0 & 1 & 0 & -1 \\
1 & 0 & 1 & 0 \\
0 & 1 & 0 & 1
\end{array}
\right)\in H_2\otimes H_3.
$$
One can easily find that rank($\widetilde{X_2})=1$. From the singular value decomposition of
$\widetilde{X_2}$, using Lemma 1 again, we get
$Y_1=\left(\begin{array}{cc}
\frac{1}{\sqrt{2}} & - \frac{1}{\sqrt{2}} \\
\frac{1}{\sqrt{2}} &  \frac{1}{\sqrt{2}}
\end{array}
\right)$, $Y_2 =I_2$, such that $X_2=Y_1\otimes Y_2$ is
unitary. That is, $(X_1\otimes Y_1 \otimes Y_2)
\rho_1(X_1\otimes Y_1 \otimes Y_2)^\dag=\rho_2$.

Our criterion is both necessary and sufficient for local equivalence of arbitrary
multipartite mixed quantum systems. However, for general degenerated states,
it could be less operational. In the following, complement to Theorem 1, we present an alternative
way to judge the local equivalence based on partial transposition of  matrices. For a density matrix $\rho \in
H_1\otimes H_2$ with entries $\rho_{m\mu,n\nu}=\langle e_m\otimes
f_\mu|\rho|e_n\otimes f_\nu\rangle$, the partial transposition of
$\rho$ is defined by \cite{MH}:
$$
\rho^{T_2}=(I\otimes
T)\rho=\sum_{mn,\mu\nu}\rho_{m\nu,n\mu}|e_m\otimes
f_{\mu}\rangle\langle e_n\otimes f_{\nu}|,
$$
where $\rho^{T_2}$ denotes the transposition of $\rho$ with respect
to the second system, $|e_n\rangle$ and $|f_\nu\rangle$ are the
bases associated with $H_1$ and $H_2$ respectively.

\noindent {\bf Theorem 2}~ Two mixed states $\rho_1$ and
$\rho_2$ in $H_1\otimes H_2$ are local unitary equivalent if and
only if $\rho_1^{T_2}$ and $\rho_2^{T_2}$ are local unitary
equivalent.

\noindent {\bf Proof}~ Without loss of generality, we assume that
$\rho_1=\sum\rho_{m\mu,n\nu}|e_m \rangle \langle
e_n|\otimes|f_\mu\rangle\langle f_\nu|$. Then
$\rho_1^{T_2}=\sum\rho_{m\nu,n\mu}|e_m \rangle \langle
e_n|\otimes|f_\mu\rangle\langle f_\nu|.$ On the one hand, if $\rho_1$
and $\rho_2$ are equivalent under local unitary transformations, one has
$$
\rho_2=(u_1\otimes u_2)\rho_1(u_1\otimes u_2)^\dag
=\sum\rho_{m\mu,n\nu}(u_1|e_m \rangle \langle
e_n|u_1^\dag)\otimes(u_2|f_\mu\rangle\langle f_\nu|u_2^\dag).
$$
Hence
$$\ba{rcl}
\rho_2^{T_2}&=&\sum\rho_{m\mu,n\nu}(u_1|e_m \rangle \langle
e_n|u_1^\dag)\otimes(u_2^\ast|f_\nu\rangle\langle f_\mu|u_2^T)\\
&=&\sum\rho_{m\mu,n\nu}(u_1\otimes u_2^\ast)(|e_m \rangle
\langle e_n|\otimes|f_\nu\rangle\langle f_\mu|)(u_1\otimes u_2^\ast)^\dag\\
&=&(u_1\otimes u_2^\ast)\rho_1^{T_2}(u_1\otimes u_2^\ast)^\dag.
\ea
$$
Therefore, $\rho_2^{T_2}$ and $\rho_1^{T_2}$ are
also local unitary equivalent.

On the other hand, since $(\rho^{T_2})^{T_2}=\rho$, if
$\rho_1^{T_2}$ and $\rho_2^{T_2}$ are equivalent under local unitary
transformations, one can derive that $\rho_1$ and $\rho_2$ are also
equivalent under local unitary transformations. \qed

The Theorem 2 is also true for $\rho^{T_1}$. Generally the partial transposed
states are no longer semi positive. They are just
Hermitian matrices. Nevertheless Theorem 2 still works for
the local unitary equivalence for Hermitian matrices.
Theorem 2 can be directly generalized to multipartite systems:

\noindent {\bf Theorem 3}~ Two mixed states $\rho_1$ and $\rho_2$ in $H_1\otimes H_2
\otimes\cdots \otimes H_n$ are local unitary equivalent if and only
if $\rho_1^{T_k}$ and $\rho_2^{T_k}$ are local unitary equivalent,
where $k\in \{1,2,...,n\}$, $\rho^{T_k}$ denotes the
transposition of $\rho$ with respect to the $k$th system.

Theorem 3 provides us an alternative way to determine the local
unitary equivalence of multipartite states. If the given states are
degenerated, the criterion given by Theorem 1 would be less
operational. In this case one may consider the partial transposition
of the states. For bipartite states, if the partially transposed
states are not degenerated, we can check the local unitary
equivalence by using the Theorem 1 and obtain the explicit local
unitary matrices. There are many degenerated states such that their
partially transposed ones are not degenerated, such as: $
\rho=\left(\begin{array}{cccc}
\frac{1}{4}& 0 & 0 & \frac{1}{16} \\
0& \frac{1}{8} & 0 & 0 \\
0& 0 & \frac{1}{8}& 0 \\
\frac{1}{16} & 0 & 0 & \frac{1}{2}
\end{array}
\right). $ If the partially transposed states are still degenerated,
but less degenerated such that they have $s$ different eigenvalues
with multiplicity $2$ and the rest eigenvalues with multiplicity
$1$, then the Theorem 1 can applied to determine the local unitary
equivalence. For the multipartite states, the Theorem 3 could be
applied to simplify problem.

{\sf Remark 3}: \ In general, implementing Theorem 1 and Theorem 3
could be with quite difficulty and complexity. Theorem 1 and Theorem
3 could be better implemented if the states are numerical or not too
complicated ones, such as the examples we given in the text, where
we can do calculations numerically by using Mathematica. But in
general, it is still an open question whether there exist an unified
algorithm to implement Theorem 1 and Theorem 3.

In summary, based on matrix realignment we have presented a
necessary and sufficient criterion of the local unitary equivalence
for general multipartite mixed quantum states, and the corresponding
explicit expression of the local unitary operators. The criterion
proposed in \cite{feij} is a special case of Theorem 1 for bipartite
case. Our criterion is even operational for a class of degenerated
states. To deal with the general degenerated states, we have also
presented another criterion based on state partial transpositions,
which, in complement to our criterion based on matrix realignment,
may transform an un-operational problem to be an operational one, so
as to make our criteria more effective. Detailed examples have been
presented. Our approach gives a new progress toward to the local
equivalence of multipartite mixed states.

\bigskip
Acknowledgments: This work is supported by the NSFC 11105226,
11275131; the Fundamental Research Funds for the Central
Universities No.12CX04079A, No.24720122013; Research Award Fund for
outstanding young scientists of Shandong Province No.BS2012DX045.


\begin{thebibliography}{99}
\bibitem{bdfk}B. Doran and F. Kirwan, Pure Appl. Math. Q {\bf 3}, 61
(2007).
\bibitem{EofCon} R. Horodecki, P. Horodecki, M. Horodecki and K. Horodecki, Rev. Mod. Phys {\bf 81}, 865 (2009).
\bibitem{nils} M.A. Nielsen and I.L. Chuang, Quantum Computation and
Quantum Information, Cambridge University Press, 2004.
\bibitem{Cbben} C.H. Bennett, D.P. Divincenzo, J.A. Smolin and W.K.
Wootters, Phys.Rev. A {\bf 54}, 3824 (1996).

\bibitem{mhor}M. Horodecki, Quant. Inf. Comput {\bf 1}, 3 (2001).

\bibitem{Auhl}A. Uhlmann, Phys. Rev. A {\bf 62}, 032307 (2000).

\bibitem{pvcm}P. Rungta, V. Buzek, C.M. Caves, M. Hillery and G. J.
Milburn, Phys. Rev. A {\bf 64}, 042315 (2001).

\bibitem{Wdur} W.~D\"ur, G. Vidal and J.I. Cirac, Phys. Rev. A {\bf
62} 062314 (2000).

\bibitem{Rains} E.M.~Rains, IEEE T. Inf. Theory {\bf 46}, 54 (2000).

\bibitem{Grassl} M.~Grassl, M.~R\"otteler and T.~Beth, Phys. Rev. A {\bf 58}, 1833 (1998).

\bibitem{makhlin} Y. Makhlin, Quant. Info. Proc {\bf 1}, 243 (2002).

\bibitem{linden} N. Linden, S. Popescu and A. Sudbery, Phys. Rev. Lett {\bf 83}, 243 (1999).

\bibitem{Linden} N. Linden and S. Popescu, Fortsch. Phys {\bf 46}, 567 (1998).

\bibitem{SFY} S. Albeverio, S.M. Fei, P.Parashar and W.L.Yang, Phys. Rev. A {\bf 68}, 010303 (2003).

\bibitem{SFG} S. Albeverio, S.M. Fei, and D.Goswami, Phys. Lett. A {\bf 340}, 37 (2005).

\bibitem{SFW} B.Z. Sun, S.M. Fei, X.Q. Li-Jost and Z.X.Wang, J. Phys. A {\bf 39}, L43-L47(2006).

\bibitem{SCFW} S. Albeverio, L. Cattaneo, S.M. Fei and X.H. Wang, Int. J. Quant. Inform {\bf 3}, 603 (2005).
\bibitem{sss} S. Szalay, J. Phys. A: Math. Theor {\bf 45} 065302 (2012)

\bibitem{mqubit} B. Kraus, Phys. Rev. Lett. {\bf 104}, 020504 (2010); Phys. Rev. A {\bf 82}, 032121 (2010).

\bibitem{bliu} B. Liu, J.L. Li, X. Li, and C.F. Qiao, Phys. Rev. Lett {\bf 108}, 050501 (2012).

\bibitem{zhou} C. Zhou, T.G. Zhang, S.M. Fei, N. Jing, and X. Li-Jost, Phys. Rev. A {\bf 86}, 010303(R) (2012).

\bibitem{zhang} T.G. Zhang, M.J. Zhao, X. Li-Jost and S.M Fei, Int. J. Theor. Phys {\bf 52}, 3020 (2013).

\bibitem{o.ru}O. Rudolph, Quantum Inf. Process {\bf4}, 219
(2005).

\bibitem{Chen}K. Chen and L.A. Wu, Quantum Inf. Comput {\bf 3}, 193
(2003).

\bibitem{AP}A. Peres, Phys. Rev. Lett {\bf 77}, 1413 (1996).

\bibitem{MH}M. Horodecki, P. Horodecki and R. Horodecki, Phys. Lett. A {\bf 223}, 1
(1996).

\bibitem{bsl}U. T. Bhosale, K. V. Shuddhodan, and A. Lakshminarayan
Phys. Rev. A {\bf 87}, 052311 (2013).

\bibitem{Raho}R.A. Horn and C.R. Johnson, Topics in matrix analysis,
Cambridge University Press, New York, (1991).

\bibitem{npp}N. P. Pitsianis, Ph.D. thesis, The Kronecker Product in
Approximation and Fast Transform Generation, Cornell University, New
York, 1997.
\bibitem{cfvl} C. F. Van Loan, J. Comput. Appl. Math {\bf 123}, 85 (2000)
\bibitem{slsx}S. Albeverio, L. Cattaneo, S.M. Fei and X.H. Wang, Int. J. Theor.
Phys {\bf 03}, 603 (2005).

\bibitem{naja}N. Courtois, A. Klimov, J. Patarin and A. Shamir, Adv.
Crypt {\bf 1807}, 392 (2000).

\bibitem{feij}S.M. Fei and N. Jing, Phys. Lett. A {\bf 342}, 77 (2005).
\end{thebibliography}
\end{document}